\documentclass[conference]{IEEEtran}
\IEEEoverridecommandlockouts

\usepackage[utf8]{inputenc}
\usepackage[english]{babel}
\usepackage{cite}
\usepackage{amsmath,amssymb,amsfonts}
\usepackage{algorithmic}
\usepackage{graphicx}
\usepackage{textcomp}
\usepackage{xcolor}
\usepackage[a4paper, total={184mm,239mm}]{geometry}
\usepackage{siunitx}
\usepackage{nicefrac}
\usepackage[acronym]{glossaries}
\usepackage[hidelinks]{hyperref}
\usepackage[capitalise]{cleveref}
\usepackage{float}
\usepackage[caption=false]{subfig}

\DeclareSIUnit\GE{GE}
\DeclareSIUnit\kGE{\kilo\GE}
\DeclareSIUnit\MGE{\mega\GE}

\begin{document}
\newcommand{\ap}{\textsc{axi-pack}}%
\newcommand{\strideSu}{5.4}%
\newcommand{\indirSu}{2.4}%
\newcommand{\strideBu}{87}%
\newcommand{\indirBu}{39}%
\newcommand{\areaOver}{6.2}%
\newcommand{\stridePowerInc}{19.8}%
\newcommand{\indirPowerInc}{15.7}%
\newcommand{\avgPowerInc}{17.8}%
\newcommand{\strideEEF}{5.3}%
\newcommand{\indirEEF}{2.1}%
\newcommand{\todo}[1]{\textit{\color{red}#1}}
\newcommand{\x}{$\times$}


\title{AXI-Pack: Near-Memory Bus Packing for Bandwidth-Efficient Irregular Workloads}

\author{
    \IEEEauthorblockN{%
    Chi Zhang\IEEEauthorrefmark{3}\textsuperscript{\textasteriskcentered}, %
    Paul Scheffler\IEEEauthorrefmark{3}\textsuperscript{\textasteriskcentered}, %
    Thomas Benz\textsuperscript{\textasteriskcentered}, %
    Matteo Perotti\textsuperscript{\textasteriskcentered}, %
    Luca Benini\textsuperscript{\textasteriskcentered}\textsuperscript{\textdagger}%
    }
    \IEEEauthorblockA{
        \textasteriskcentered~\textit{Integrated Systems Laboratory, ETH Zurich}, Switzerland \\
        \textdagger~\textit{Department of Electrical, Electronic, and Information Engineering, University of Bologna}, Italy \\
        \{chizhang,paulsc,tbenz,mperotti,lbenini\}@iis.ee.ethz.ch
    }
    \thanks{%
        \IEEEauthorrefmark{3} Both authors contributed equally to this research.
    }
}


\maketitle

\begin{abstract}
Data-intensive applications involving irregular memory streams are inefficiently handled by modern processors and memory systems highly optimized for regular, contiguous data.
Recent work tackles these inefficiencies in hardware through core-side stream extensions or memory-side prefetchers and accelerators, but fails to provide end-to-end solutions which also achieve high efficiency in on-chip interconnects.
We propose AXI-Pack, an extension to ARM's AXI4 protocol introducing bandwidth-efficient strided and indirect bursts to enable end-to-end irregular streams.
AXI-Pack adds irregular stream semantics to memory requests and avoids inefficient narrow-bus transfers by packing multiple narrow data elements onto a wide bus. It retains full compatibility with AXI4 and does not require modifications to non-burst-reshaping interconnect IPs.
To demonstrate our approach end-to-end, we extend an open-source RISC-V vector processor to leverage AXI-Pack at its memory interface for strided and indexed accesses. On the memory side, we design a banked memory controller efficiently handling AXI-Pack requests.
On a system with a 256-bit-wide interconnect running FP32 workloads, AXI-Pack achieves near-ideal peak on-chip bus utilizations of \strideBu\% and \indirBu\%, speedups of {\strideSu}x and {\indirSu}x, and energy efficiency improvements of {\strideEEF}x and {\indirEEF}x over a baseline using an AXI4 bus on strided and indirect benchmarks, respectively.
\end{abstract}

\newacronym{hpc}{HPC}{high performance computing}
\newacronym{ml}{ML}{machine learning}
\newacronym{isa}{ISA}{instruction set architecture}
\newacronym{fp}{FP}{floating-point}
\newacronym{dl}{DL}{deep learning}
\newacronym{la}{LA}{linear algebra}
\newacronym{ip}{IP}{intellectual property}
\newacronym[firstplural=systems-on-chip (SoCs)]{soc}{SoC}{system-on-chip}

\newacronym{mac}{MAC}{multiply-accumulate}
\newacronym{fem}{FEM}{finite element analysis}
\newacronym{simd}{SIMD}{single-instruction, multiple-data}
\newacronym{rtl}{RTL}{register transfer level}
\newacronym{dlt}{DLT}{data layout transform}

\newacronym{fifo}{FIFO}{first in, first out}
\newacronym{fu}{FU}{functional unit}
\newacronym{alu}{ALU}{arithmetic logic unit}
\newacronym{fpu}{FPU}{floating-point unit}
\newacronym{ssr}{SSR}{stream semantic register}
\newacronym{issr}{ISSR}{indirection stream semantic register}
\newacronym{tcdm}{TCDM}{tightly-coupled data memory}
\newacronym{dma}{DMA}{direct memory access}
\newacronym{sm}{SM}{streaming multiprocessor}
\newacronym{vlsu}{VLSU}{vector load-store unit}

\newacronym{spvv}{SpVV}{sparse vector-vector multiplication}
\newacronym{spmv}{SpMV}{sparse vector-matrix multiplication}
\newacronym{spmm}{SpMM}{sparse matrix-matrix multiplication}
\newacronym{csrmv}{CsrMV}{CSR matrix-vector multiplication}
\newacronym{csrmm}{CsrMM}{CSR matrix-matrix multiplication}

\newacronym{csf}{CSF}{compressed sparse fiber}
\newacronym{csr}{CSR}{compressed sparse rows}
\newacronym{csc}{CSC}{compressed sparse columns}
\newacronym{bcsr}{BCSR}{blocked compressed sparse rows}

\newacronym{axi4}{AXI4}{Advanced eXtensible Interface 4}
\newacronym{amba}{AMBA}{Advanced Microcontroller Bus Architecture}
\newacronym{sram}{SRAM}{static random-access memory}



\section{Introduction}
\label{sec:intro}

Growing performance demands and large, sparse datasets in domains like machine learning\cite{Hoefler2021SparsityID}, graph analytics\cite{Min2020EMOGI}, fluid dynamics\cite{Georgescu2010ConjugateGO}, and recommender systems\cite{Li2017CuSNMFAS} push data-driven applications toward increasingly irregular data access patterns.
This poses a challenge to general-purpose CPUs~\cite{Talati2021ProdigyIT,Wang2019StreambasedMA} and GPUs~\cite{MndezLojo2012AGI,Li2017CuSNMFAS} optimized for highly regular compute.
To keep their functional units highly utilized and achieve satisfactory performance and energy efficiency, \gls{simd} architectures require contiguous data chunks not naturally found in irregular workloads.  
Memory hierarchies are also tuned to contiguous, high-locality data and struggle with irregular access patterns\cite{Talati2021ProdigyIT,Wang2019StreambasedMA}, resulting in long latencies, poor bandwidth utilization, and cache thrashing.

Existing research aims to improve irregular workload performance and tackle these shortcomings through \emph{core-side} or \emph{memory-side} hardware extensions. 
Core-side extensions often use \emph{stream} abstractions \cite{Talati2021ProdigyIT, Wang2019StreambasedMA,Schuiki2021StreamSR, Scheffler2021IndirectionSS,Domingos2021UnlimitedVE} to describe entire sequences of irregular accesses, freeing processors from address calculation and decoupling memory accesses from execution. 
Most works focus on accelerating \emph{strided} and \emph{indirect} streams, which are most common in practice \cite{Wang2019StreambasedMA}.
Mapping these streams to architectural registers \cite{Wang2019StreambasedMA,Schuiki2021StreamSR, Scheffler2021IndirectionSS,Domingos2021UnlimitedVE} further improves functional unit utilization and enables significant speedups.
However, these works largely ignore downstream interconnects and memory systems. While some authors propose high-level cache policies to avoid thrashing \cite{Talati2021ProdigyIT, Wang2019StreambasedMA}, they do not address fundamental limitations like the high index-fetching overhead of core-side indirection and the inherent inefficiency of narrow bus accesses in address-based interconnects.

In contrast, memory-side extensions prefetch and accelerate irregular accesses using pattern-aware memory controllers \cite{Carter1999ImpulseBA}, prefetchers \cite{Hussain2017ANH, Hussain2018MemoryCF}, and \gls{dlt} accelerators \cite{Barredo2021PLANARAP,Lloyd2015InMemoryDR}. Unlike core-side extensions, these solutions reduce access times and prevent narrow bus accesses by \emph{packing} fetched irregular elements into bus-wide lines, which are then mapped to virtual addresses \cite{Carter1999ImpulseBA}, written to internal scratchpads \cite{Hussain2017ANH, Hussain2018MemoryCF}, or written back to memory \cite{Barredo2021PLANARAP,Lloyd2015InMemoryDR}. However, these solutions have their own drawbacks: they occupy virtual or physical memory and lack the tight architectural integration of core-side extensions, limiting their acceleration potential and complicating programming.

Thus, while existing core- and memory-side extensions tackle inefficiencies in their respective domains, they forego each other's benefits and do not integrate with established on-chip interconnect protocols, failing to provide an  end-to-end solution for bandwidth-efficient irregular streams.

To address these shortcomings, we propose \ap, an extension to Arm's widespread \gls{axi4} on-chip protocol enabling end-to-end, tightly-packed strided and indirect memory streams. 
\ap~transparently extends \gls{axi4}'s existing contiguous bursts, leveraging their decoupled, latency-resilient nature. 
It remains compatible with all existing \gls{axi4} features and even existing interconnect blocks that do not reshape bursts. It encodes stream semantics (stride or index base and size) directly into burst requests, ensuring performance and flexibility even for short streams. Indirection is efficiently handled at memory endpoints. 
In principle, \ap~supports non-core requestors (e.g., accelerators) and systems with multiple requestors and endpoints.

To demonstrate \ap~in an end-to-end full-system context, we extend an open-source RISC-V vector processor for efficient strided and indexed accesses and design a banked memory controller efficiently handling irregular bursts.
On a system with a 256-bit-wide \ap~bus running various irregular FP32 workloads, we achieve bus utilizations of up to \SI{\strideBu}{\percent} on strided and \SI{\indirBu}{\percent} on indirect benchmarks, resulting in peak speedups of \strideSu\x~and \indirSu\x~over a baseline with a standard \gls{axi4} bus.
We implement our evaluation system in GlobalFoundries' 22nm FD-SOI technology and find that {\ap} improves energy efficiency by up to {\strideEEF}\x~and {\indirEEF}\x~in strided and indirect benchmarks while incurring only  \SI{\areaOver}{\percent} of our vector processor's area for our controller.
Finally, we analyze the impact of element and index size as well as bank count on \ap~performance and controller complexity.

To summarize, our contributions are as follows:
\begin{enumerate}
    \item We extend the widespread high-performance on-chip protocol \gls{axi4} to support end-to-end bus-packed strided and indirect streams with full backward compatibility.
    \item We extend an open-source RISC-V vector processor with {\ap} to enable high bus efficiencies and significant speedups on irregular workloads, and demonstrate a simple banked memory controller to serve irregular bursts.
    \item We evaluate {\ap} by benchmarking irregular workloads on our extended vector processor, achieving bus utilizations of up to \SI{\strideBu}{\percent} and \SI{\indirBu}{\percent} and speedups of up to \strideSu\x~and \indirSu\x~for strided and indirect workloads.
    \item We evaluate our {\ap} system and controller in terms of timing, area, and energy efficiency benefits, finding energy efficiency improvements of up to {\strideEEF}\x~and {\indirEEF}\x.
\end{enumerate}

\section{Architecture}
\label{sec:arch}

\subsection{AXI-Pack Protocol}

\begin{figure}[t!]
  \centering
  \includegraphics[width=0.92\linewidth]{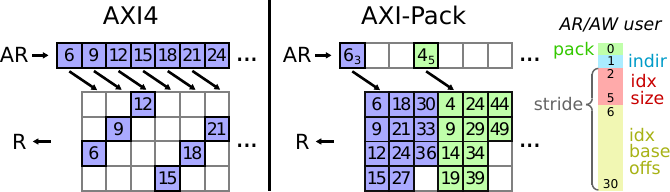}
  \caption{\ap~AR/AW user extensions and strided read example}
  \label{fig:ap_user}
\end{figure}

\ap~extends Arm's \gls{axi4} \cite{Arm2021AMBAAXI}, a widely-adopted high-performance non-coherent on-chip memory protocol. \gls{axi4} defines five independent channels: AR and AW carry read and write requests, R and W carry read and write data, and B carries the write response. Without extensions, linear, fixed, and wrapping bursts are supported. Each channel provisions a user field of parametric width that allows extending functionality without compromising compatibility with the baseline protocol.

\ap~extends the request channels AR and AW with user signals to support packed irregular bursts as illustrated in \Cref{fig:ap_user}. The \texttt{pack} bit indicates whether our extension is used, while the \texttt{indir} bit differentiates between strided and indirect bursts. The remaining bits are shared between both burst types; they indicate either the element stride for strided bursts or the index size and base offset for indirect bursts. 

While active, the new irregular burst types alter the semantics of existing channel fields. Most notably, data elements of the requested size, scattered in memory, are tightly packed onto the R and W data buses to fully utilize them. Additionally, the start of irregular bursts is aligned with the bus instead of the address to simplify feeding data to and from vectorized functional units. Finally, the AR and AW \texttt{size} fields, usually only changed for narrow beats, indicate the data element size.

In addition to performance and bus utilization, these semantics aim to maximize the \emph{transparency} and \emph{portability} of \ap: they ensure that any existing \gls{axi4} \gls{ip} blocks handling non-modifiable transactions without splitting, such as the routing blocks provided in \cite{Kurth2022AnOP}, are already compatible with \ap~without any modifications. \glspl{ip} that require burst splitting or reshaping, such as bus width converters, can easily be extended to support \ap~by re-packing bus-aligned data elements as for existing burst types.

\subsection{Vector Processor Extension}

To demonstrate the benefits and flexibility of \ap, we extend the open-source \emph{Ara} \cite{new_ara} RISC-V vector processor to leverage it for efficient irregular memory accesses. Ara acts as a co-processor to the CVA6 core \cite{Zaruba2019TheCO},  which dispatches vector instructions to Ara. Both access memory over \gls{axi4}. 

As mandated by its instruction set, Ara supports three vector memory access types: \emph{contiguous}, \emph{strided}, and \emph{indexed}. Without extensions, only contiguous accesses can leverage bursts. For strided and indexed accesses, Ara must compute the address and issue individual narrow accesses for each element, leaving the data channels severely underutilized as shown in \Cref{fig:ap_user}.

\Cref{fig:ara_extension} shows our extensions to Ara. We modify its \gls{vlsu} to use \ap~for strided and indexed vector accesses. For strided accesses, we simply translate the existing \texttt{vlse} and \texttt{vsse} instructions to \ap~requests and exchange the read or written data directly with vector registers or functional lanes for chaining.
The existing indexed access instructions \texttt{vl(o|u)xei} and \texttt{vs(o|u)xei} in the RISC-V vector extension presume that indices are already loaded into vector registers, necessitating the move of indices into the core and precluding efficient memory-side indirection. To remedy this, we extend Ara's decoder and introduce two new \emph{in-memory} indexed access instructions, \texttt{vlimxei} and \texttt{vsimxei}, which use index arrays in memory for indirection. These instructions can directly be translated to indirect \ap~bursts and allow for packed bus data to be exchanged with registers or lanes without format changes.

\subsection{Banked Memory Controller}

\begin{figure*}[t!]
  \subfloat[Vector processor extensions]{
    \label{fig:ara_extension}
    \includegraphics[width=0.23\linewidth, height=0.3\linewidth]{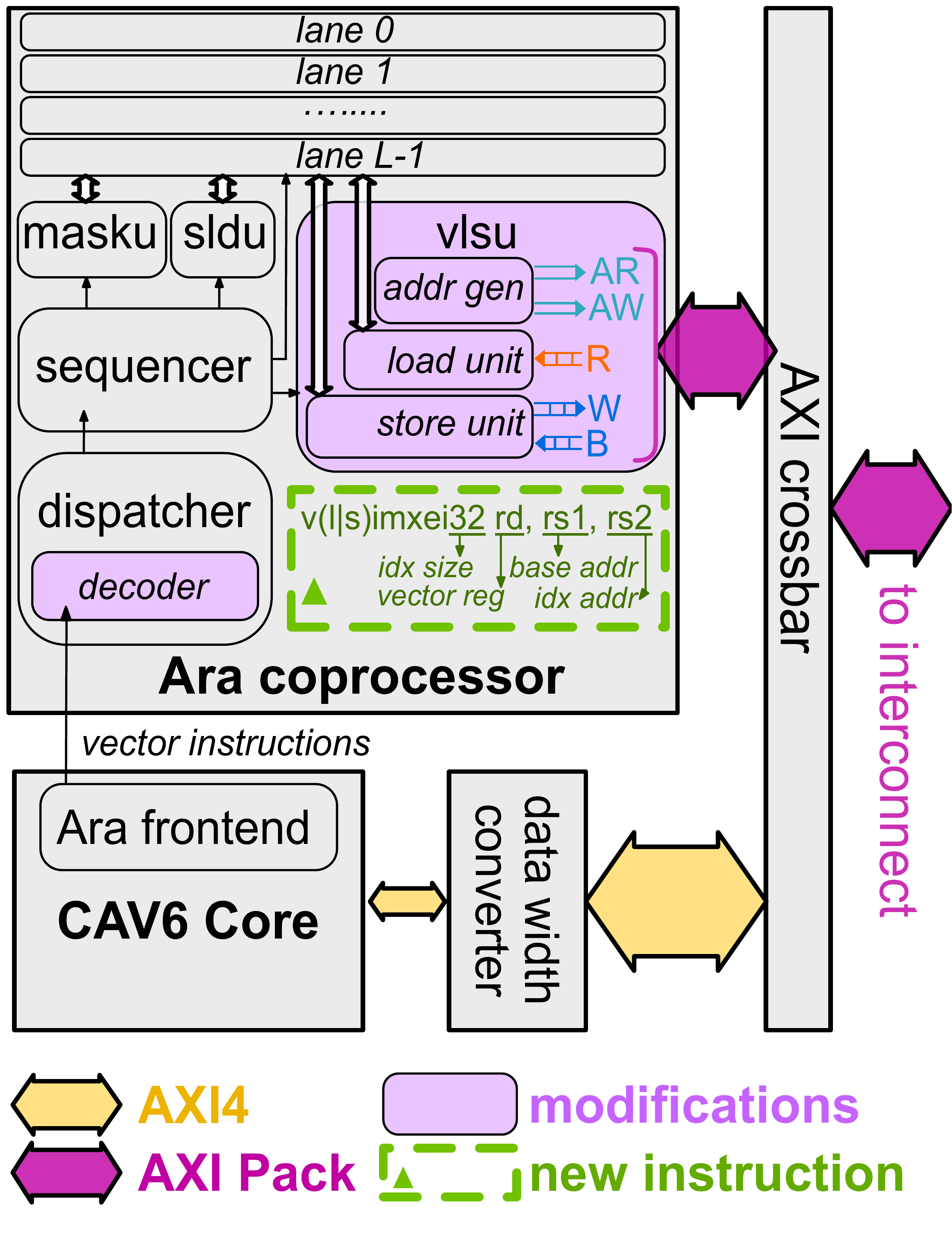}
  }
  \subfloat[Controller top level]{
    \label{fig:arch_contr_top}
    \includegraphics[width=0.22\linewidth, height=0.3\linewidth]{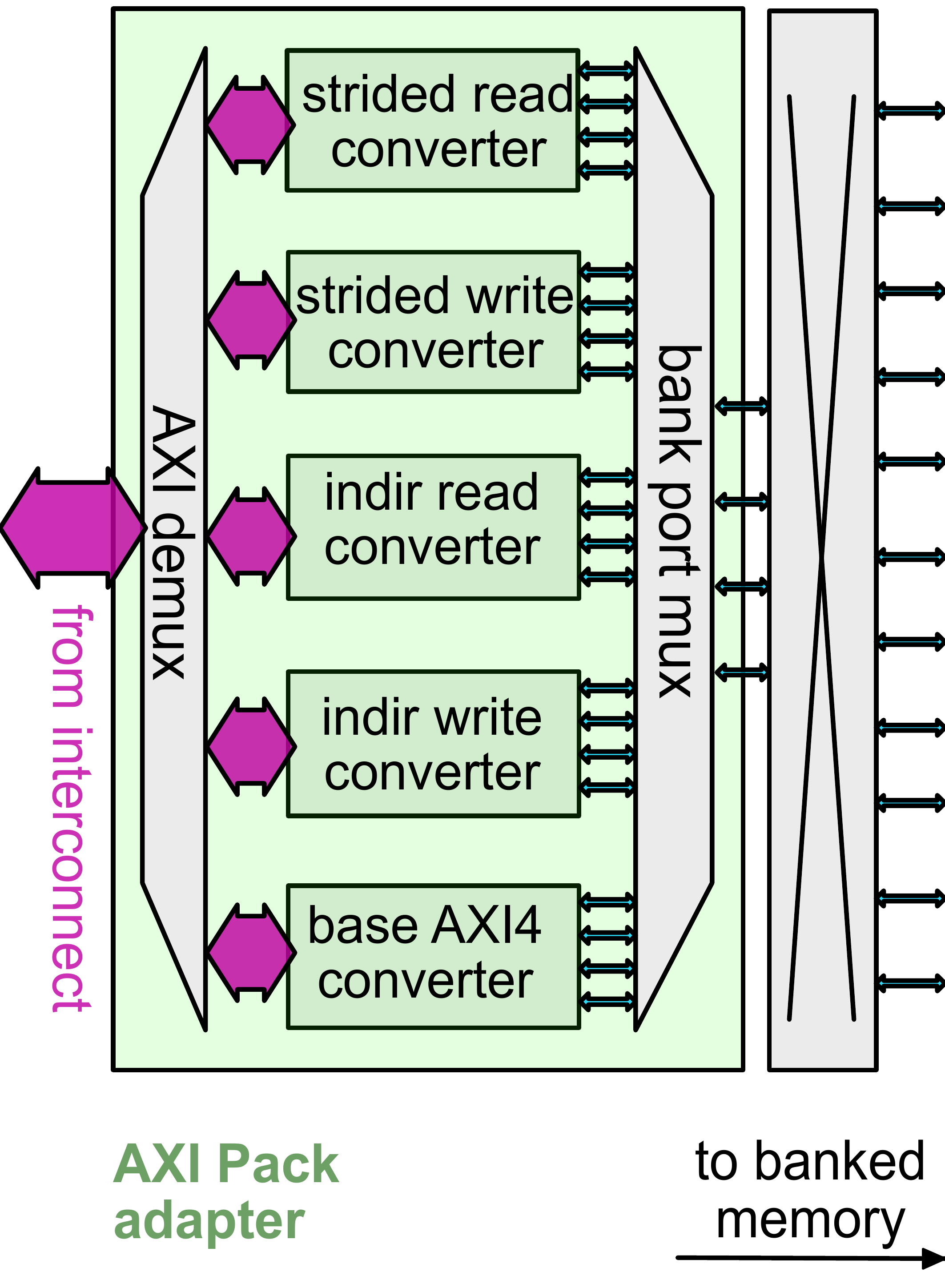}
  }
  \subfloat[Strided read converter]{
    \label{fig:arch_contr_rstrided}
    \includegraphics[width=0.2223\linewidth, height=0.3\linewidth]{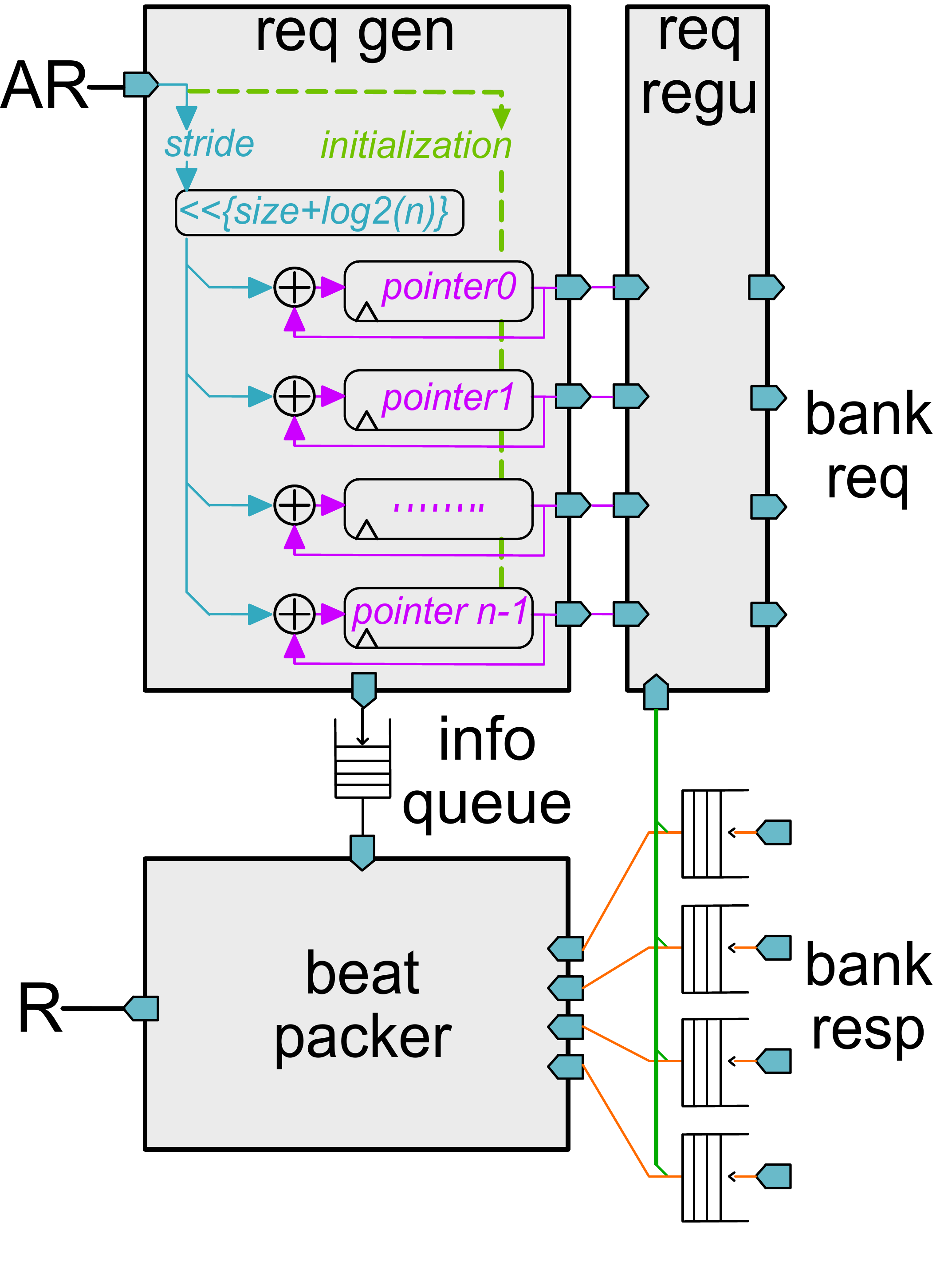}
  }
  \subfloat[Indirect read converter] {
    \label{fig:arch_contr_rindir}
    \includegraphics[width=0.2446\linewidth, height=0.3\linewidth]{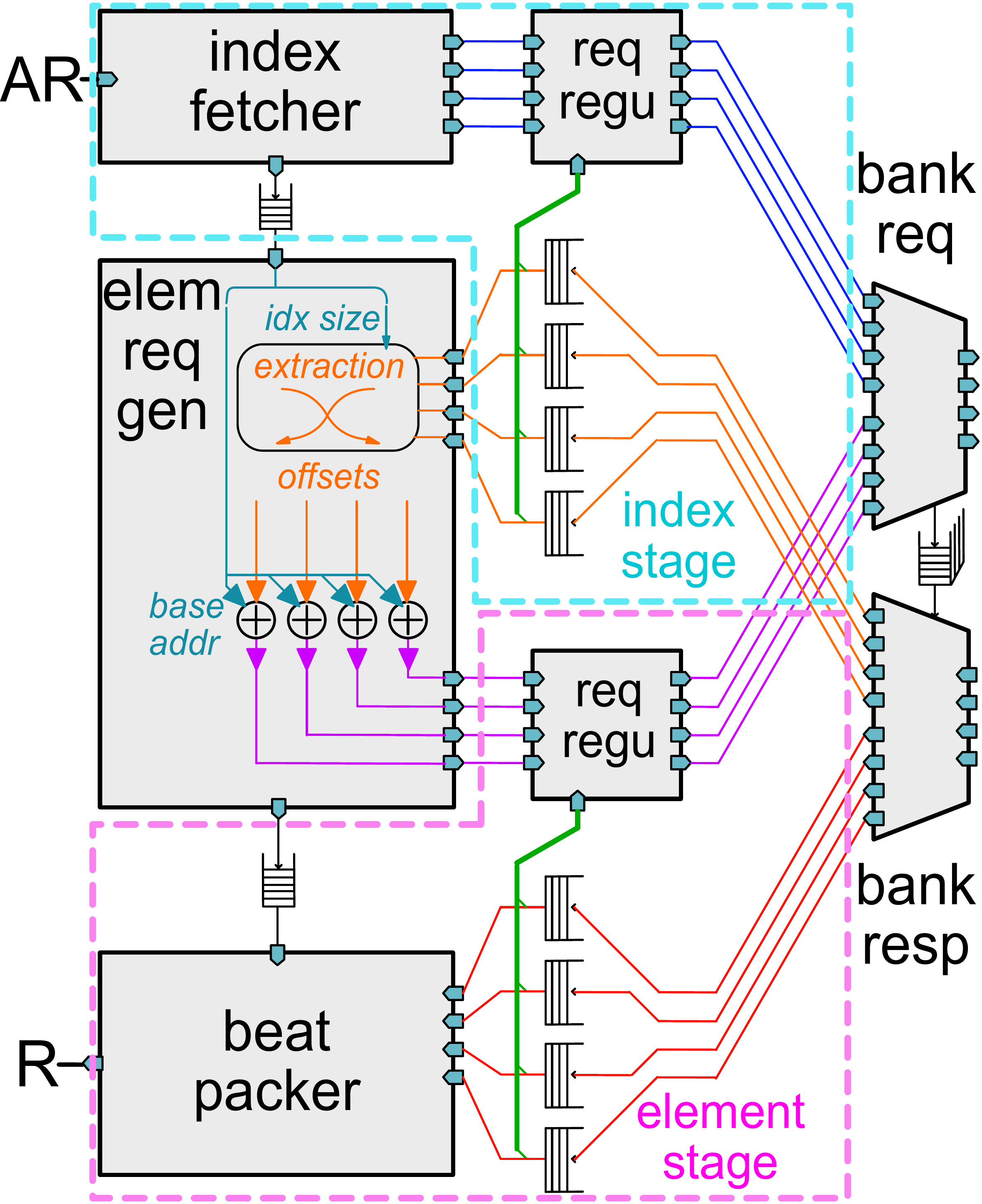}
    }
  \caption{\ap~processor extension, multi-banked controller, and converter architecture}
  \label{fig:ap_contr}
\end{figure*}

To demonstrate the efficient handling of \ap~at banked memory endpoints, we design a proof-of-concept controller translating \ap~requests to sequences of parallel banked memory accesses. Our controller is fully backward-compatible with and efficiently handles regular \gls{axi4} bursts.

\Cref{fig:arch_contr_top} shows the controller architecture. The \emph{adapter} translates both regular and irregular bursts to sequences of $n$ parallel \emph{word} accesses, where a word is the same width $W$ as the used memory banks and determines the smallest efficiently-handled element size. For $D$-bit-wide \ap~data buses, $n=D/W$, since we must read or write $D/W$ words in parallel to saturate them. The adapter connects to an $n\times m$ crossbar mapping the $n$ word access ports to $m$ interleaved banks.

Internally, the adapter forwards requests to one of five converters which may concurrently handle bursts. The \emph{base} converter handles regular \gls{axi4} bursts, while the remaining converters are dedicated to strided and indirect read and write operations, respectively. Handling reads and writes individually leverages the inherent concurrency of the R and W channels.

\Cref{fig:arch_contr_rstrided} details the strided read converter architecture. For each beat in a burst, the \emph{request generator} issues $n$ parallel word requests fetching the elements to be packed and pushes metadata needed for later packing into an \emph{info} queue. The words read from the banks are stored in decoupling queues and then passed to the \emph{beat packer}, which packs the words as specified by metadata popped from the \emph{info} queue to form the R beats. To prevent word queue overflows, a \emph{request regulator} limits the number of requests in-flight for each word lane.

\Cref{fig:arch_contr_rindir} shows the indirect read converter architecture. It involves two stages sharing the $n$ word request ports through round-robin arbitration: the \emph{index stage} fetches indices from memory and the \emph{element stage} uses these indices to fetch indirect elements and pack them into R beats. The index stage is analogous to the strided read converter, but issues only contiguous word requests. The fetched indices are passed to the \emph{element request generator}, which shifts and adds them to the specified base address to generate word requests for the desired elements. Finally, the requested elements are packed by a beat packer as specified by metadata from the element request generator to form the desired R beats.

The corresponding strided and indirect \emph{write} converters are similar and differ only in the direction of the datapath: a \emph{beat unpacker} splits beats into individual words, which are then used as write data for the generated write requests. The memory responses are combined and forwarded to the B channel.

\section{Evaluation}
\label{sec:eval}

\subsection{Setup and Workloads}

To evaluate \ap, we consider three RISC-V \glspl{soc} using CVA6 with Ara as a vector processor, \gls{axi4} interconnects, and a banked on-chip \glsunset{sram}\gls{sram} memory:

\begin{itemize}
    \item \textsc{base}: unmodified CVA6 and Ara connecting to a regular banked memory over a standard \gls{axi4} bus.
    \item \textsc{pack}: unmodified CVA6 and \ap-extended Ara connecting to a banked memory with an \ap~controller over an \ap-extended bus.
    \item \textsc{ideal}: like \textsc{base}, but Ara connects directly to an exclusive, idealized memory with one port per lane, serving data with ideal packing, bandwidth, and latency.
\end{itemize}

\textsc{ideal} provides an upper bound for possible \ap~benefits by idealizing interactions between Ara's \gls{vlsu} and memory. However, it does not avoid  inefficiencies arising from Ara's internal microarchitecture or CVA6.
In all systems, Ara is parameterized to eight vector lanes and 256-bit-wide data buses. The banked memories provide eight 32-bit-wide word ports backed by 17 banks, which we determine in \cref{sec:eval_parsens} to provide a good area-performance tradeoff.

On each system, we evaluate a set of vectorized benchmarks benefiting from efficient strided and indirect memory accesses:

\begin{itemize}
    \item \texttt{ismt}: \emph{in-situ matrix transpose}. We transpose a square matrix in place by swapping and rotating elements above and below the diagonal using strided accesses.
    
    \item \texttt{gemv}: \emph{general matrix-vector multiply}. We investigate both \emph{row-} and \emph{column-wise} dataflows, with the latter trading vector reductions for strided accesses, and use the fastest approach on each system for fair comparisons.
    
    \item \texttt{trmv}: \emph{triangular matrix-vector multiply}, a \texttt{gemv} with an upper-triangular matrix. Only nonzero elements are streamed, incurring bursts of varying lengths. We again use the fastest dataflow on each system.
    
    \item \texttt{spmv}: \emph{sparse matrix-vector multiply}, a widespread irregular memory-bound operation using indirect accesses.
    
    \item \texttt{prank}: \emph{PageRank} \cite{Page1999ThePC}, which rates each node in a graph based on the edges inbound to it. The graph is represented as a sparse weighted adjacency matrix.
    
    \item \texttt{sssp}: \emph{single-source shortest path}, which calculates the shortest path from one node to all others in a weighted, directed graph represented as a sparse matrix.
\end{itemize}

We run the first three benchmarks leveraging strided streams on randomly-generated square matrices and the latter three leveraging indirect streams on real-world sparse matrices from the SuiteSparse collection \cite{Davis2011TheUO} in the widespread \gls{csr} format. Elements are stored as 32-bit floats and indices as 32-bit integers. On indirect workloads, the \textsc{pack} system uses our extensions to handle indirection in-memory, whereas \textsc{base} and \textsc{ideal} systems fetch indices into Ara.

\subsection{Performance}
\label{subsec:res_perf}

\begin{figure}[t!]
\centering
  \subfloat[Speedups and bus utilizations across workloads]{
    \label{fig:all-bench-perf}
    \includegraphics[width=0.94\linewidth, height=0.354\linewidth]{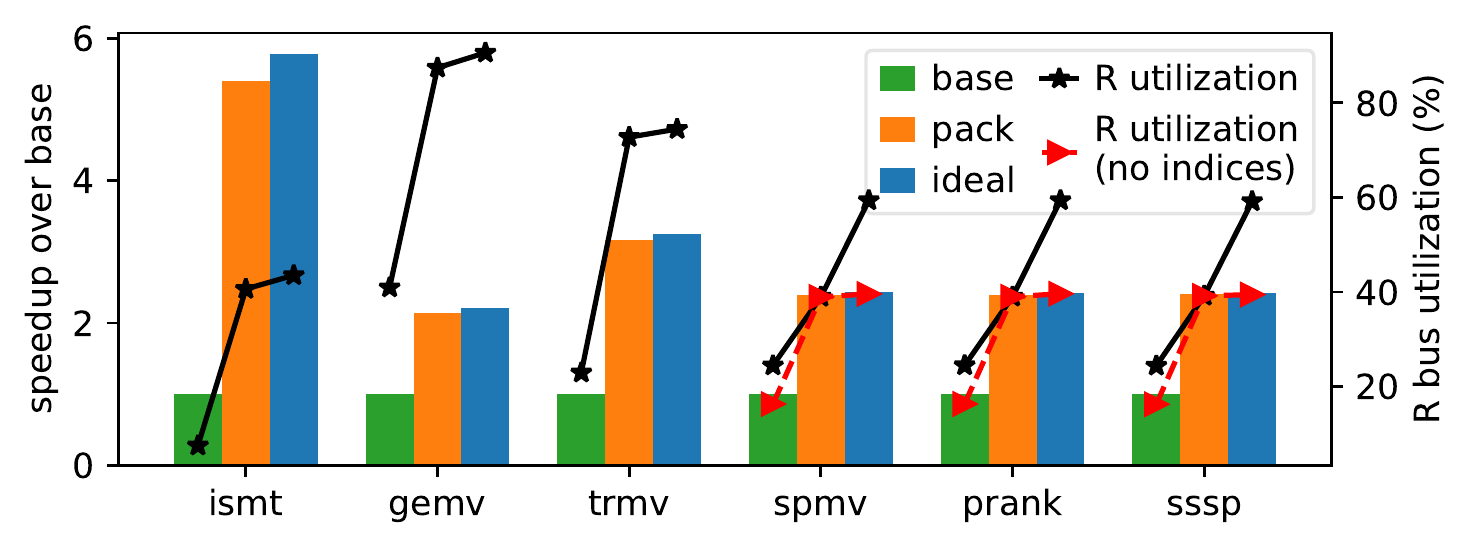}
  }
  
  \subfloat[\texttt{gemv} dataflows compared]{
    \label{fig:gemv-comp}
    \includegraphics[width=0.45\linewidth, height=0.34\linewidth]{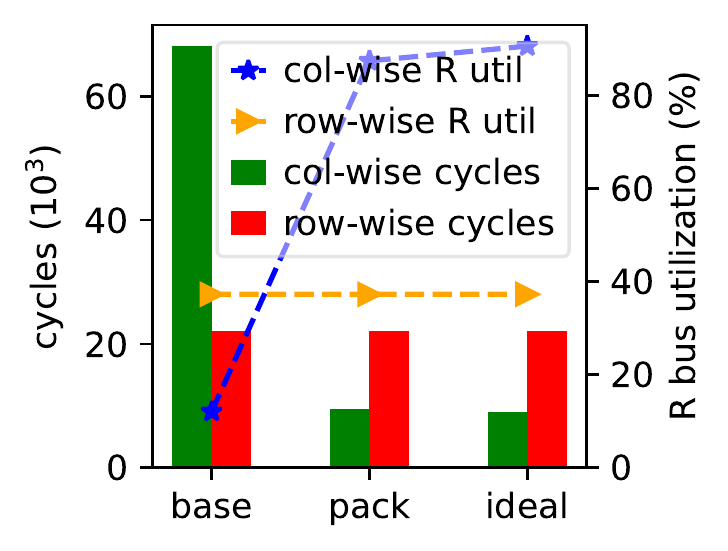}
  }
  \subfloat[\texttt{trmv} dataflows compared] {
    \label{fig:trmv-comp}
    \includegraphics[width=0.45\linewidth, height=0.34\linewidth]{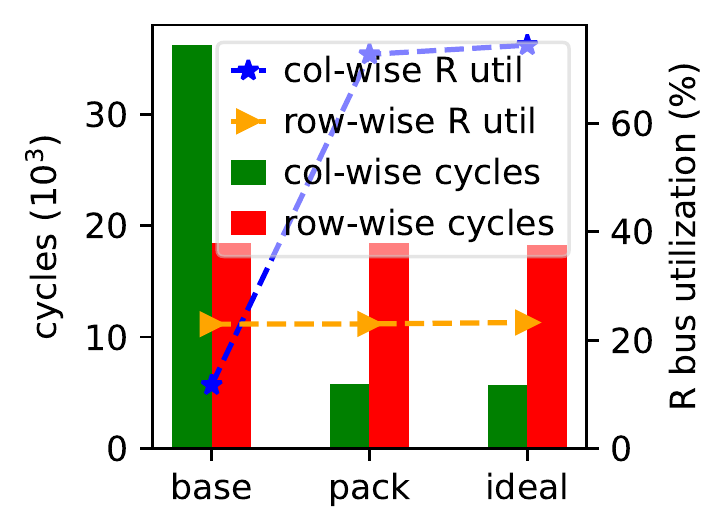}
    }
    
  \subfloat[\texttt{ismt} \textsc{pack} speedup scaling]{
    \label{fig:ismt-speedup}
    \includegraphics[width=0.45\linewidth, height=0.31\linewidth]{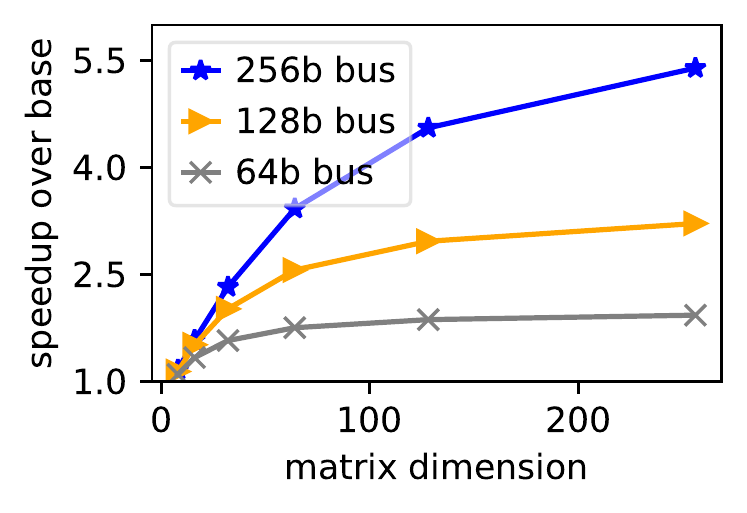}
  }
  \subfloat[\texttt{spmv} \textsc{pack} speedup scaling] {
    \label{fig:spmv-speedup}
    \includegraphics[width=0.45\linewidth, height=0.31\linewidth]{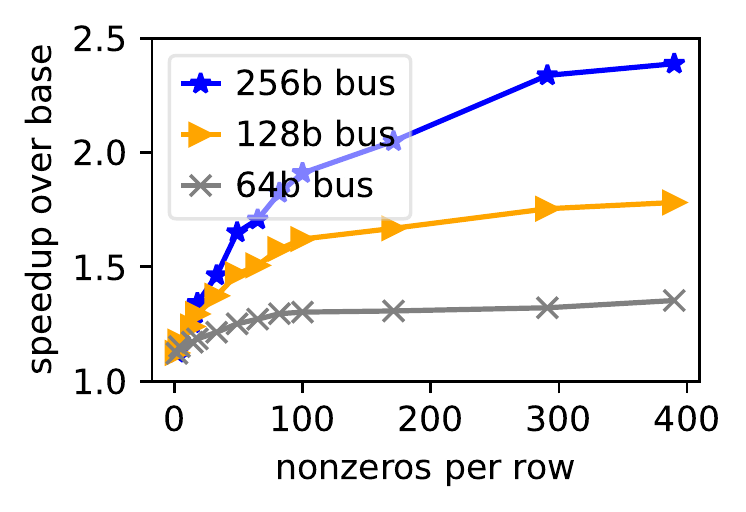}
    }
  \caption{\ap~performance results}
  \label{fig:ap_perf}
\end{figure}

We simulate our systems at the \gls{rtl} to determine the performance and read bus utilization for each benchmark.
We initially assume a fixed matrix size of 256 for strided workloads and the sparse matrix \texttt{heart1} (390 average nonzeros per row) for indirect workloads.
\texttt{gemv} and \texttt{trmv} use a column-wise dataflow on \textsc{pack} and \textsc{ideal} and a row-wise dataflow on \textsc{base}, which we will show to be optimal for each respective system.

\cref{fig:all-bench-perf} shows \textsc{pack} speedups over \textsc{base} and read bus utilizations with and without index transfers. \ap~significantly improves bus utilization and performance for all workloads, achieving {\SI{97}{\percent}} of the \textsc{ideal} performance on average.
On strided workloads, we measure peak speedups of \strideSu\x~(\texttt{ismt}) and bus utilizations of  \SI{\strideBu}{\percent} (\texttt{gemv}). We note that read bus utilizations on \texttt{ismt} are limited to \SI{50}{\percent} due to read-write ordering in Ara.
On indirect workloads, we achieve speedups of up to \indirSu\x~({\texttt{spmv}}) and bus utilizations of up to \SI{\indirBu}{\percent} ({\texttt{sssp}}). \textsc{pack} handles indirection directly in its \ap~controller, avoiding \textsc{ideal}'s waste of up to {\SI{20}{\percent}} (\texttt{spmv}) of bus time on index traffic and shifting indexed workloads further from the memory-bound toward the compute-bound regime.

\cref{fig:gemv-comp,fig:trmv-comp} compare the row- and column-wise dataflow performance for \texttt{gemv} and \texttt{trmv}. Row-wise flows use only long contiguous accesses, so their performance is identical for \textsc{base} and \textsc{pack} and very close to \textsc{ideal}. However, they require costly vector reductions, limiting \textsc{base} bus utilizations to \SI{37}{\percent} and \SI{23}{\percent}. Column-wise flows avoid reduction by working on multiple results at once, providing higher \textsc{ideal} performance and \textsc{pack} utilizations of \SI{87}{\percent} and \SI{72}{\percent}. However for \textsc{base}, we stick to a row-wise flow, as the performance impact of strided accesses outweighs that of reductions without our extensions.

We also analyze the impact of input size and bus width on \ap~speedups for representative strided and indirect workloads. 
\Cref{fig:ismt-speedup} shows \texttt{ismt} speedups for matrix dimensions of 8 to 256 and bus widths of 64 to 256 bit, corresponding to 2 to 8 Ara lanes. 
As matrix size increases, speedups converge and reach up to 1.9, 3.2, and 5.4\x; as we widen the bus, the narrow accesses of \textsc{base} become less efficient, increasing peak \textsc{pack} speedups. 
As matrix size decreases, streams and useful computation phases shorten and become bottlenecked by the overhead of row iteration, decreasing speedups.
\Cref{fig:spmv-speedup} shows \texttt{spmv} speedups for sparse matrices with 2 to 390 average nonzeros per row and the same bus widths as before.
Speedups again converge and reach up to 1.4, 1.8, and 2.4\x. We see similar scaling trends as for \texttt{ismt} because in \texttt{spmv}, the nonzeros per row determine the computation phase and stream lengths in each row iteration. We note that thanks to our request-bundling approach, using \ap~never results in a slowdown no matter how short streams become.

\subsection{Area and Timing}

We synthesize our {\ap} adapter with Synopsys \emph{Design Compiler} for GlobalFoundries' \SI{22}{\nano\meter} FD-SOI technology, targeting the SSG corner at \SI{-40}{\celsius} with low-$V_t$ cells, 0.72V supply voltage, and no back-biasing. Unless otherwise specified, we constrain a \SI{1}{\giga\hertz} clock and \SI{100}{\pico\second} IO delays and parameterize the decoupling queues to a depth of four.

\begin{figure}[t!]
\centering
  \subfloat[Adapter area versus minimum clock]{
    \label{fig:a_t}
    \includegraphics[width=0.543\linewidth, height=0.375\linewidth]{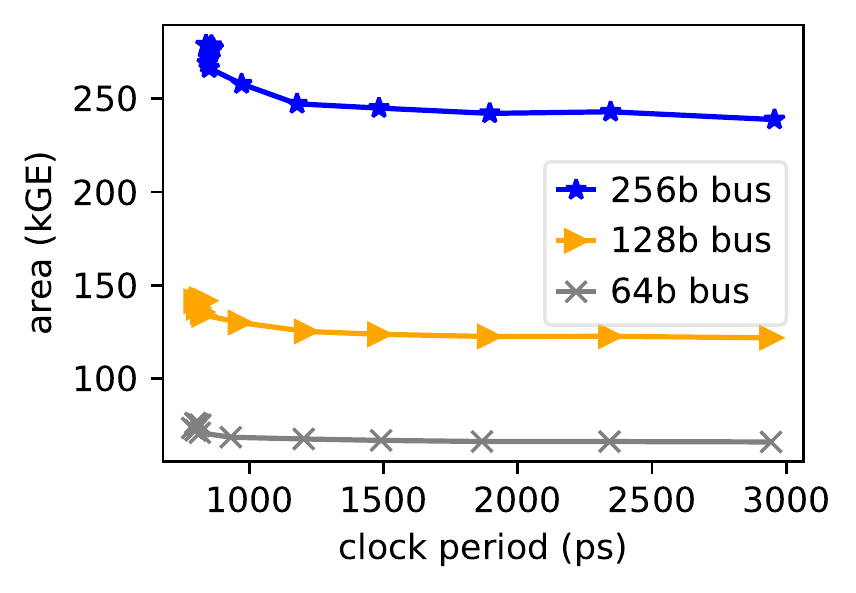}
  }
  \subfloat[Adapter area breakdown] {
    \label{fig:area_break}
    \includegraphics[width=0.356\linewidth, height=0.375\linewidth]{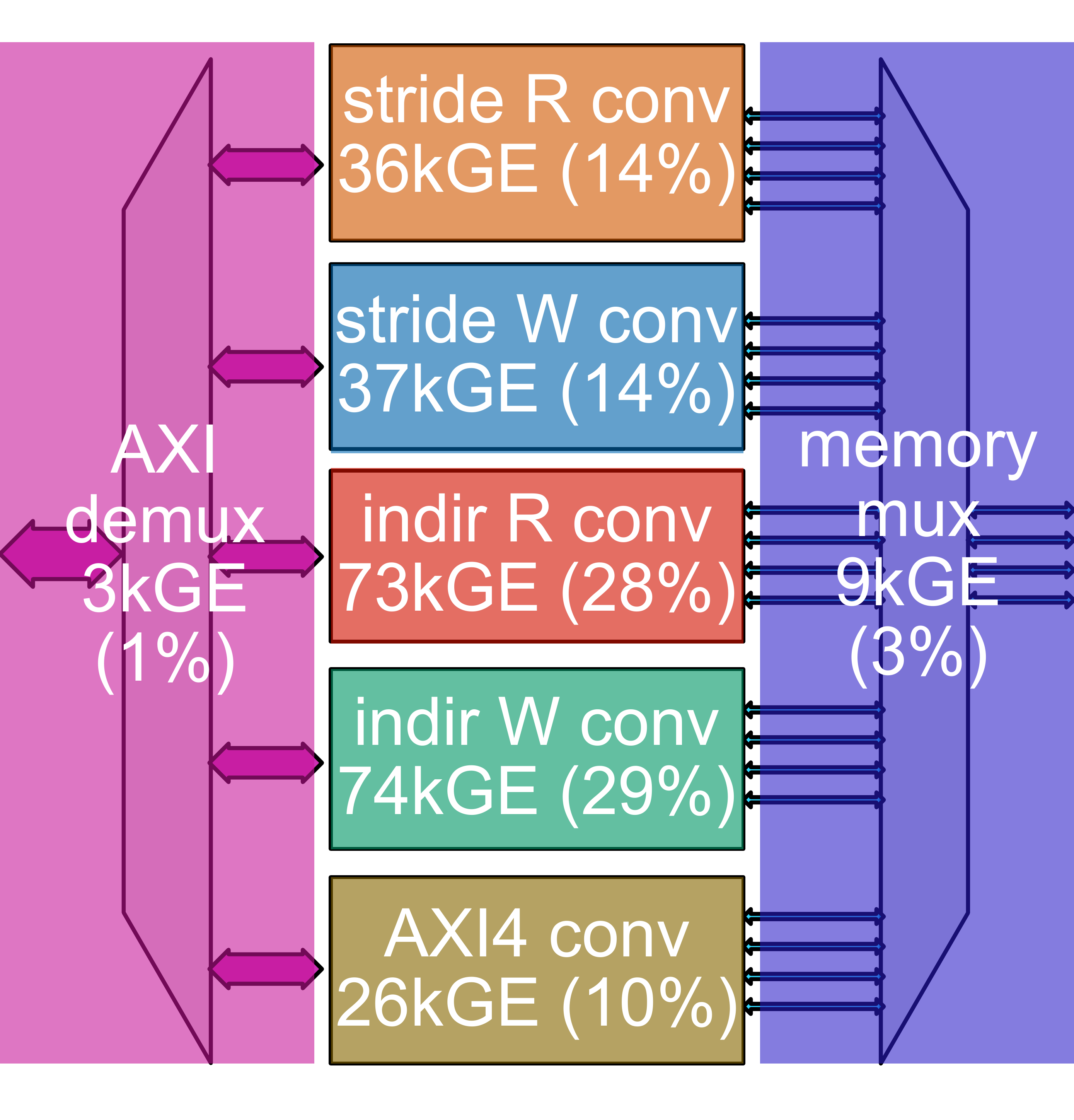}
    }
    
  \subfloat[Benchmark powers and energy efficiency improvements]{
    \label{fig:power_analysis}
    \includegraphics[width=0.96\linewidth, height=0.356\linewidth]{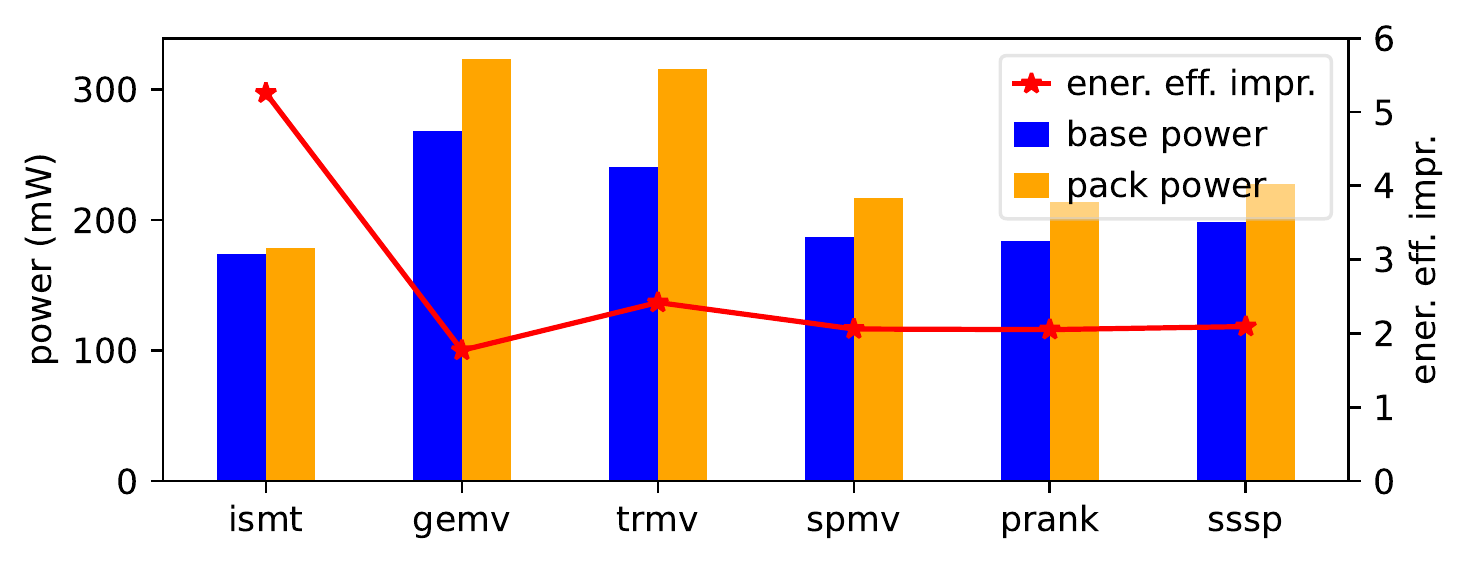}
  }
  \caption{\ap~ area, timing, and energy results}
  \label{fig:ap_synthesis}
\end{figure}

\cref{fig:a_t} shows the minimum achieved clock period and area for different clock constraints and bus widths of 64, 128, and 256 bit. Our adapter shows good scalability, increasing linearly in area with bus width and incurring 69, 130 and \SI{257}{\kGE} at \SI{1}{\giga\hertz}. Our full 256-bit controller incurs merely \SI{\areaOver}{\percent} of Ara's area, demonstrating that \ap~handling at banked endpoints is reasonably inexpensive.
As we decrease the constrained clock, we see that adapter area scales gracefully past Ara's \SI{1}{\giga\hertz} clock target and reaches minimum periods of 787, 800, and \SI{839}{\pico\second} with only small increases in area.

\cref{fig:area_break} shows a hierarchical area breakdown of the adapter. As expected, the read and write converters are similar in size for both irregular burst types, since they simply reverse each other's datapaths. While the simpler strided converters are only up to \SI{42}{\percent} larger than the base \gls{axi4} converter, the indirect converters are nearly double this size due their two stages.

\subsection{Energy and Power}

We estimate the power consumption of \textsc{pack} and \textsc{base}, excluding the SRAM crossbar and banks, in the TT corner of GlobalFoundries' \SI{22}{\nano\meter} FD-SOI technology at \SI{1}{\giga\hertz}. We topographically synthesize our system using Synopsys \emph{Design Compiler} and estimate power on the benchmarks from \Cref{subsec:res_perf} using Synopsys \emph{PrimeTime}. \Cref{fig:power_analysis} shows the average power and energy efficiency improvement of \textsc{pack} over \textsc{base} for each benchmark. Despite small power increases in \textsc{pack} by at most \SI{31}{\percent} (\texttt{trmv}), all workloads see notable energy efficiency improvements, achieving peaks of {\strideEEF}\x~(\texttt{ismt}) and {\indirEEF}\x~(\texttt{sssp}) on strided and indirect benchmarks, respectively.

\subsection{Parameter Sensitivity}
\label{sec:eval_parsens}

To gain deeper insight into the scaling of \ap~performance and hardware complexity, we investigate the impact of element and index size as well as bank count on read bus utilization and bank crossbar area. 
For performance measurements, we connect our controller to an ideal requestor issuing continuous read requests of length 256 and use random indices.
Unless otherwise specified, parameters default on their \textsc{pack} system configuration, but we increase decoupling queue depths to 32 to avoid bottlenecks unrelated to our analysis. 
We consider power-of-two bank counts from 8 to 32, which result in minimal addressing logic, as well as prime counts in this range, which minimize bank conflicts across different strides. We also consider an ideal memory without bank conflicts.

\begin{figure}[t!]
\centering

  \subfloat[Indirect read utilization] {
    \label{fig:indirect_snes}
    \includegraphics[width=0.96\linewidth, height=0.48\linewidth]{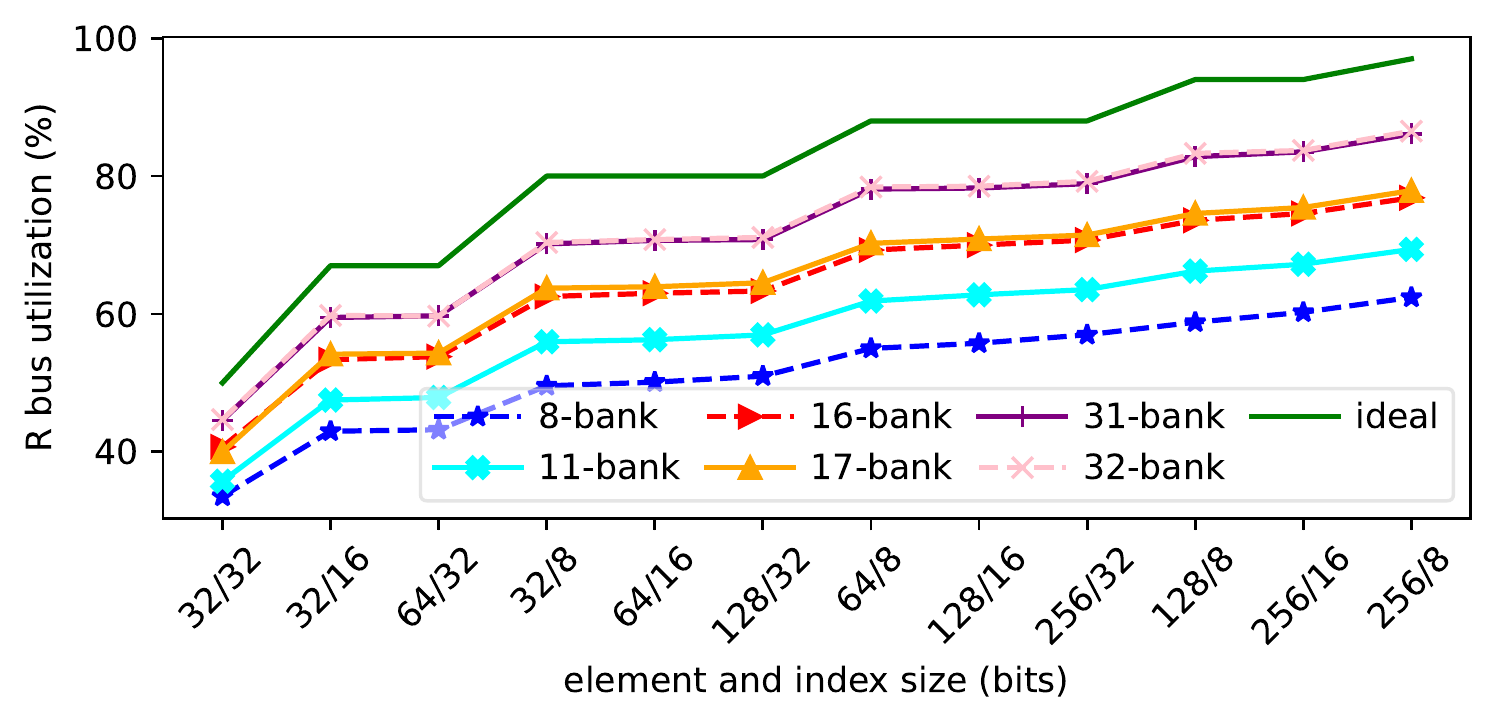}
    }
    
  \subfloat[Strided read utilization]{
    \label{fig:stride_snes}
    \includegraphics[width=0.45\linewidth, height=0.3375\linewidth]{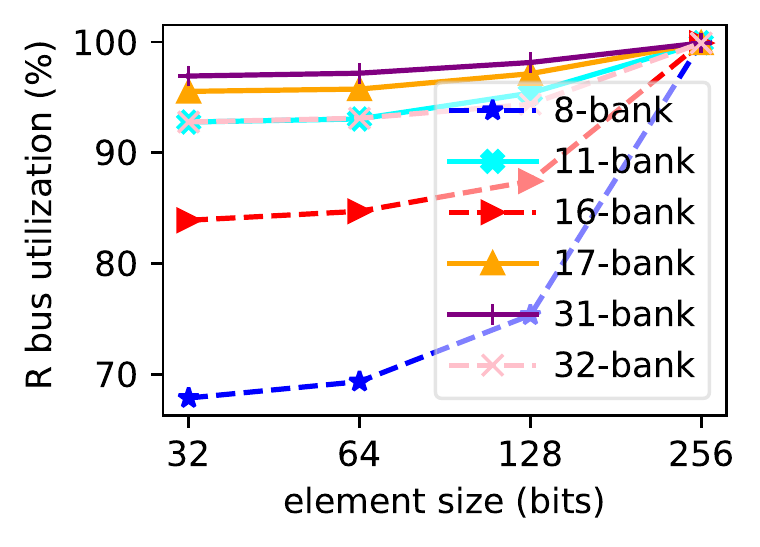}
  }
  \subfloat[Bank crossbar area]{
    \label{fig:area_tcdm}
    \includegraphics[width=0.45\linewidth, height=0.3375\linewidth]{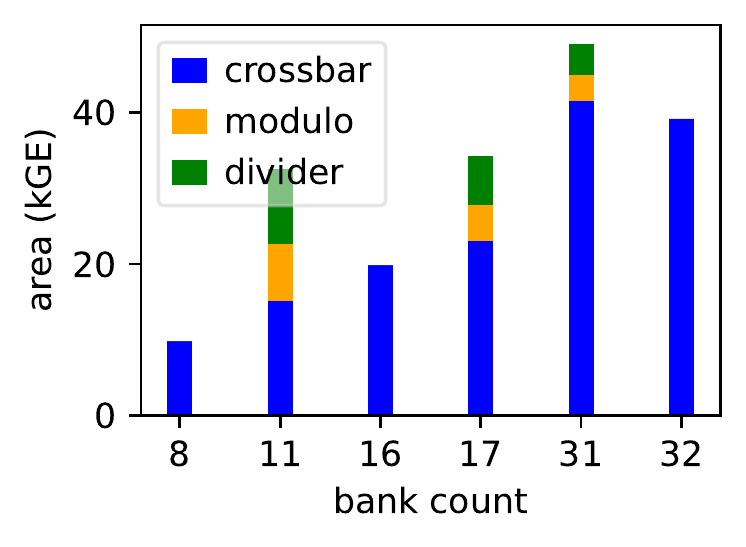}
  }

  \caption{\ap~parameter sensitivity results}
  \label{fig:ap_sens}
\end{figure}

\paragraph*{Indirect accesses} \Cref{fig:indirect_snes} shows the bus utilization achieved on indirect reads for different element-index size pairs and bank counts. For all size pairs, utilization increases monotonically with bank count as fewer bank conflicts occur. Since indirect bursts involve one contiguous and one random but no strided bank access sequences, prime bank counts show no inherent advantage here. 
Across all bank counts, utilization improves mainly with the \emph{ratio} $r$ of element size to index size: since we fetch indices as whole bus lines, we must fetch one index line for every $r$ data beats on average, limiting our ideal bus utilization to $\nicefrac{r}{r+1}$. For 32-bit elements and index sizes of 32, 16, and 8 bit, this corresponds to ideal utilizations of 50, 67, and \SI{80}{\percent}. Thus, with larger elements or smaller indices, \ap~indirection bus utilizations can further exceed those shown in \cref{subsec:res_perf}.

\paragraph*{Strided accesses} \Cref{fig:stride_snes} shows the bus utilization for strided reads for different element sizes and bank counts, averaged across element strides of 0 to 63. As expected, prime bank counts offer significant performance benefits on strided accesses, though more banks further improve performance for both power-of-two and prime bank counts. With increasing element size, conflicts become less likely for all bank counts as there are fewer aligned elements in each bus-wide line.

\paragraph*{Bank crossbar area} \Cref{fig:area_tcdm} shows the bank crossbar's total area for different bank counts, highlighting the overhead prime bank counts incur for modulo and division units to compute bank addresses. Power-of-two-banked crossbars are generally cheaper and prime-banked overheads decrease with increasing bank counts. Since 17 banks provide good area-overhead and area-performance tradeoffs (95\% and 81\% of ideal performance on strided and indirect reads on average), this is the bank count we chose for our evaluation systems.

\section{Related Work}
\label{sec:relwork}

Existing hardware approaches to efficient strided and indirect streams focus mostly on either end of the memory system. 

\emph{Core-side extensions} decouple accesses from execution and eliminate redundant load-store and bookkeeping instructions.
Prodigy \cite{Talati2021ProdigyIT} prefetches nested indirect streams and proposes dynamic cache bypass policies. While highly decoupled, it does not simplify program flow, limiting its acceleration. 
Wang et al. \cite{Wang2019StreambasedMA} propose strided and indirect streams mapped directly to architectural registers. This eliminates load-store and address iteration instructions, but still incurs dedicated instructions to step streams.
Stream semantic registers \cite{Schuiki2021StreamSR} and their indirection extensions \cite{Scheffler2021IndirectionSS} implicitly step streams on access, enabling near-continuous useful instruction issues even on single-issue in-order cores. Domingos et al. \cite{Domingos2021UnlimitedVE} extend register-mapped irregular streams to vector processors. 
Except for cache policies, these extensions ignore interconnect and memory system efficiency. \ap~is largely orthogonal to all of them; it may be used with any bus width, burst length, or mapping mechanism, providing a reusable protocol carrying irregular streams through interconnects and to stream-aware endpoints with high bus efficiency.

\emph{Memory-side extensions} focus on bus efficiency and access latency. 
The Impulse memory controller \cite{Carter1999ImpulseBA} maps irregular streams to virtual pages; it provides inherent, on-the-fly bus packing, but relies on managed virtual addressing.
Hussain et al. propose pattern-aware memory controllers \cite{Hussain2018MemoryCF} and systems \cite{Hussain2017ANH} prefetching irregular stream descriptors to dedicated scratchpads. This enables fast, packed core accesses, but incurs notable complexity overheads.
PLANAR \cite{Barredo2021PLANARAP} accelerates layout transforms by writing packed, cacheable irregular data to memory ahead of use, and the data rearrangement engine \cite{Lloyd2015InMemoryDR} is integrated directly into a hybrid memory cube architecture. While \gls{dlt} accelerators are highly bandwidth-efficient, they require physical memory buffers and explicit, ahead-of-time invocation to be beneficial.
\ap~enables the benefits of all of the above extensions. Bus packing can be done on the fly by our controller or ahead of time by an \ap-capable \gls{dma} controller. Our lightweight irregular requests provide performance without precluding the use of more complex, memory-mapped stream descriptors. However unlike other proposals, \ap~builds on an established protocol and extends irregular streams throughout interconnects, feeding directly into cores in an end-to-end fashion.

\section{Conclusion}
\label{sec:concl}

In this work, we present {\ap}, an extension to the widespread \gls{axi4} on-chip protocol enabling highly-efficient end-to-end strided and indirect memory streams. \ap~is fully backward-compatible and enables decoupled, bus-packed streams whose semantics are directly encoded into requests, ensuring high performance and flexibility even for short streams.
To demonstrate \ap~in an end-to-end system, we extend an open-source RISC-V vector processor to use it for strided and indexed accesses and design a banked memory controller serving irregular bursts. We evaluate the performance of the resulting vector processor system by evaluating benchmarks involving strided and indirect accesses.
{\ap} increases bus utilizations up to {\strideBu}\% in strided and {\indirBu}\% in indirect benchmarks, resulting in speedups of up to {\strideSu}\x~and {\indirSu}\x. Synthesizing our {\ap} controller, we find that it incurs only {\areaOver}\% of the area of Ara, but improves energy efficiency by up to {\strideEEF}\x~in strided and {\indirEEF}\x~in indirect workloads.


\bibliographystyle{IEEEtran}
\bibliography{IEEEabrv,main}

\end{document}